\documentclass[12pt]{article}
\usepackage{amssymb}
\usepackage{epsfig}

\catcode`\@=11
\def\numberbysection{\@addtoreset{equation}{section}
        \def\theequation{\thesection.\arabic{equation}}}

\def\beq{\begin{equation}}
\def\eeq{\end{equation}}
\numberbysection
\begin{document}
\begin{titlepage}
\begin{center}
\hfill  \\
\vskip 1.in {\Large \bf Path integral and noncommutative poisson brackets} \vskip 0.5in P. Valtancoli
\\[.2in]
{\em Dipartimento di Fisica, Polo Scientifico Universit\'a di Firenze \\
and INFN, Sezione di Firenze (Italy)\\
Via G. Sansone 1, 50019 Sesto Fiorentino, Italy}
\end{center}
\vskip .5in
\begin{abstract}
We find that in presence of noncommutative poisson brackets the relation between Lagrangian and Hamiltonian is modified. We discuss this property by using the path integral formalism for non-relativistic systems. We apply this procedure to the harmonic oscillator with a minimal length.
\end{abstract}
\medskip
\end{titlepage}
\pagenumbering{arabic}
\section{Introduction}

The lack of a consistent theory of quantum gravity has inspired many proposals to overcome this problem. Between them we have explored in the last few years the possibility, suggested by the high energy limit of string theory \cite{1}, of deforming the canonical commutation rules giving rise to a non-zero minimal uncertainty in position measurements. This introduces a natural $UV$ cutoff from the beginning, which is also compatible with Lorentz invariance. It is pretty clear that this eventual minimal observable length should occur at Planck scale.

Unfortunately there are very few results obtained in this direction, and we felt that is was necessary to derive a more systematic approach. In this paper we have studied the interconnection between the path integral formalism and the non-commutative Poisson brackets which prepare the classical system to the non-commutative quantization ( see also \cite{2}).

For technical reasons we have confined our study to the $1d$ non-relativistic case where well defined results can be obtained, postponing the generalization to the Snyder geometry to a future research.

We carefully analyze all the steps which usually transform the transition amplitude to a functional integration, finding that, at least in the $1d$ non-relativistic case, the main difference with the canonical quantization can be characterized with a modification of the relation between the Lagrangian and the Hamiltonian.
Already at a classical level the same equations of motion, obtained with the non-commutative Poisson brackets
in the Hamiltonian formalism, can be derived with the modified Lagrangian found in the path integral  approach.

To test our formal derivation on a practical example, we successfully apply the non-commutative path integral formalism to the harmonic oscillator recovering a known correspondence between the harmonic oscillator with a minimal length and an ordinary problem of quantum mechanics.

\section{Deformed quantum mechanics}

In the $1d$ non-relativistic case, we are going to substitute the classical Poisson brackets with noncommutative ones defined as

\beq \{ x, p \} \ = \ 1 \ + \ \beta p^2 \label{21} \eeq

The corresponding noncommutative operatorial quantization is formally obtained by promoting the phase space coordinates to linear operators acting in a linear space $ \cal{H} $ together with the condition that these operators
satisfy to the commutation rules given by $ i \hbar $ times the corresponding noncommutative Poisson brackets

\beq [ \hat{x}, \hat{p} ] \ = \ i \hbar ( 1 \ + \ \beta p^2 ) \label{22} \eeq

Given a generic state $ | \psi > $ in the linear space $ \cal{H} $ its temporal evolution is determined by the Hamiltonian operator

\beq i \hbar \ \frac{\partial}{\partial t} \ | \psi > \ = \ \hat{H} \ | \psi > \label{23} \eeq

Given an initial state $ | \psi_i > $, describing a system at an initial time $ t_i $, the solution of the Schrodinger equation is formally given by, for time independent Hamiltonians,

\beq | \psi (t) > \ = \ e^{-\frac{i}{\hbar} \ \hat{H} \ ( t - t_i )} \ | \psi_i > \label{24} \eeq

Physically we are interested in the transition amplitude that the system is described at a final time $ t_f $
by the state described by $ | \psi_f > $, which is obtained by projecting on this state the solution of the Schrodinger equation

\beq < \psi_f | \psi (t_f) > \ = \ < \psi_f | e^{-\frac{i}{\hbar} \ \hat{H} \ ( t_f - t_i )} \ | \psi_i > \label{25} \eeq

In this article we will concentrate on obtaining representations of this amplitude through the functional integrals.

\section{Functional integral in phase space}

We will specify the Hamiltonian operator for a particle with mass $ m $ living in one spatial dimension as

\beq H( \hat{x}, \hat{p} ) \ = \ \frac{p^2}{2 m} \ + \ V(x) \label{31} \eeq

Without loss of generality we can consider the matrix element

\beq A \ = \ < p_f | e^{-\frac{i}{\hbar} \hat{H} T } | p_i > \label{32} \eeq

 where $ T = ( t_f - t_i ) $ is the time of particle propagation and $ p_i, p_f $ are eigenstates of the momentum operator. We can use the completeness relation for the momentum eigenstates

\beq I = \int dp \ | p >< p | \ \ \ \ \ \ < p | p' > \ = \ \delta( p - p') \label{33} \eeq

The functional integral connected to the non commutative Poisson brackets (\ref{21}) can be derived in the following way.
We can represent the transition amplitude (\ref{32}) as a product of $ N $ factors and then we can insert $ N-1 $ times
the completeness relation

\begin{eqnarray} A & = & < p_f |{( e^{ -\frac{i T }{\hbar N} \hat{H}} )}^N | p_i > \ = \
 < p_f |  \ e^{ -\frac{i \epsilon }{\hbar} \hat{H}} \ e^{ -\frac{i \epsilon }{\hbar} \hat{H}} \ ... \ e^{ -\frac{i \epsilon }{\hbar} \hat{H}} \ | p_i > \nonumber \\
 & = & \int ( \prod^{N-1}_{k=1} \ d p_k ) \ \prod^{N}_{k=1} \ < p_k | \ e^{ -\frac{i \epsilon }{\hbar} \hat{H}} \ | p_{k-1} > \label{34} \end{eqnarray}

where we have defined $ p_0 = p_i $, $ p_N = p_f $, $ \epsilon = \frac{T}{N} $.

The next step is using $ N $ times the completeness relation for the coordinate eigenstates ( studied in \cite{3})

\beq I = \int \frac{dx}{\gamma} \ | x >< x | \ \ \ \ \ \ < x | x' > \ \neq \ \delta( x - x') \label{35}\eeq

This frame is not orthonormal due to the space fuzziness. Indeed the eigenstates $ | x > $ form a overcomplete basis because they contain a one parameter family of orthonormal frames ( for details see \cite{3} )). We end up with the following formula

\beq A \ = \ \int ( \prod^{N-1}_{k=1} \ d p_k ) \ \int ( \prod^{N}_{k=1} \ \frac{dx_k}{\gamma} ) \
\prod^{N}_{k=1} \ < p_k | x_k > \ < x_k | \ e^{ -\frac{i \epsilon }{\hbar} \hat{H}} \ | p_{k-1} > \label{36} \eeq

This is still an exact formula, but now we are going to use approximations which are well defined only in the limit
$ N \rightarrow \infty $ ( $ \epsilon \rightarrow 0 ) $.  To derive the functional integral it is necessary to evaluate the following matrix element

\beq < x | e^{ -\frac{i \epsilon }{\hbar} \hat{H}( \hat{x}, \hat{p} ) } | p > \ = \ < x | p > e^{ -\frac{i \epsilon }{\hbar} H( x, p ) } \label{37} \eeq

where $ H(x,p) = \frac{p^2}{2m} + V(x) $.

A crucial point for the noncommutative case is substituting for the wave function of the coordinate eigenstates  ( derived in \cite{3} )

\beq < p | x > \ \sim \ e^{ -\frac{i p \cdot x }{\sqrt{\beta p^2}}  \arctan \sqrt{\beta p^2 }   } \label{38} \eeq

At this level the transition amplitude is represented by integrals that do not contain operators anymore. It is now easy recognizing that in the $ \epsilon \rightarrow 0 $ limit the exponent reduces to the following classical action

\begin{eqnarray} A & = & \int {\cal D} p \ {\cal D} x \ e^{ -\frac{i}{\hbar}  S[x,p] }
 \nonumber \\
 S[x,p] & = & - \int^T_0 \ dt \ \left( \frac{\dot{p}}{ 1 + \beta p^2 } \ x \ + \ H(x,p) \right)
 \label{39} \end{eqnarray}

 where $ T = N \epsilon $ is the total propagation time. In general this amplitude is not else that the formal sum on all possible paths weighted by the exponent of  $ \frac{i}{\hbar} $ times the classical action.

Therefore we have derived the form of the Lagrangian corresponding to the non-commutative Poisson brackets (\ref{21}) which can be recast also as

 \begin{eqnarray} {\cal L} & = &
 \frac{\arctan \sqrt{\beta p^2 }}{\sqrt{\beta}} \ \dot{x} \ - \ H(x,p)
 \nonumber \\ \{ x, p \} & = & 1 + \beta p^2
 \label{310} \end{eqnarray}

The non-commutative deformation has strongly modified the relation between the Lagrangian and the Hamiltonian for a generic $1d$ non-relativistic system.

 \section{Classical solutions}

The equations of motion for a generic Hamiltonian $ H(x,p) $ generated by the modified Poisson brackets (\ref{21}) are of the type ( see also \cite{4} )

 \begin{eqnarray} \dot{x} & = & \{ x, H \}
 \nonumber \\ \dot{p} & = & \{ p, H \}
 \label{41} \end{eqnarray}

Remarkably they are still compatible with energy conservation

 \beq E \ = \ \frac{p^2}{2m} + V(x) \label{42} \eeq

from which it is possible to derive the following first integral

 \beq \int^x_{x(0)} \ \frac{dx}{\sqrt{2m ( E - V(x)) } ( 1 \ + \ \beta \ 2m ( E - V(x)) } \ = \ \frac{t}{m} \label{43} \eeq

Specializing to the harmonic potential

 \beq V(x) = \frac{1}{2} \ m \omega^2 x^2 \label{44} \eeq

one can derive by using energy conservation

 \beq \dot{x} \ = \ \frac{p}{m} \ ( 1 + \beta p^2 ) \ = \ \sqrt{\frac{2E}{m}} \ ( 1 + \beta \ (2 m E) ) \
 \sqrt{ 1 - a x^2 } \ ( 1 - b x^2 ) \label{45} \eeq

 where the parameters

 \beq a = \frac{m \omega^2}{2E} \ \ \ \ \ \ \ b = \frac{\beta m^2 \omega^2}{1 + \beta ( 2 m E ) } \label{46} \eeq

The first integral (\ref{43}) is solved in this case by

 \begin{eqnarray} x & = & \frac{ \sin \xi }{ \sqrt{ a - b \ cos^2 \xi } } \ \ \ \ \ \ \ \xi = \omega t \sqrt{ 1 +
  \beta \ (2 m E)  } \nonumber \\
 p & = & \sqrt{ 2 m E } \frac{ \sqrt{ a - b } \ \cos \xi }{ \sqrt{ a - b \ cos^2 \xi } }
 \label{47} \end{eqnarray}

It is worth noticing that the same equations of motion can be described with a Lagrangian formalism

 \beq {\cal L}_{ H O } \ = \ \frac{\arctan \sqrt{\beta p^2 }}{\sqrt{\beta}} \ \dot{x} \ - \ \frac{p^2}{2m} \ - \
 \frac{1}{2} \ m \omega^2 \ x^2
 \label{48} \eeq

where the relation between $p$ and $\dot{x}$ is implicitly defined by

 \beq \dot{x} =  \frac{p}{m} \ ( 1 + \beta p^2 ) \label{49} \eeq

Now we are ready to quantize the harmonic oscillator with non-commutative Poisson brackets, known as the harmonic oscillator with a minimal length.

\section{Quantization of the harmonic oscillator}

By using the results of the third section, the non-commutative quantization of the harmonic oscillator can be rigorously defined by the following functional integral

\begin{eqnarray} A & = & \int {\cal D} p \ {\cal D} x \ e^{ -\frac{i}{\hbar}  S_{H O} [x,p] }
 \nonumber \\
 S_{H O} [x,p] & = & - \int^T_0 \ dt \ \left( \frac{\dot{p}}{ 1 + \beta p^2 } \ x \ + \ \frac{p^2}{2m} \ + \ \frac{1}{2} \ m \omega^2 \ x^2 \right)
 \label{51} \end{eqnarray}

Let us notice that, thanks to the particular potential of the harmonic oscillator, the functional integration in
$x$ space is simply a Gaussian integral and can be solved exactly

\begin{eqnarray} A & = & \int {\cal D} p \ e^{ -\frac{i}{\hbar}  \tilde{S} [p] }
 \nonumber \\
 \tilde{S} [p] & = &  \int^T_0 \ dt \ \left( \frac{\dot{p}^2}{ 2 m \omega^2 \ ( 1 + \beta p^2 )^2 }  \ - \ \frac{p^2}{2m} \right)
 \label{52} \end{eqnarray}

The remaining integral can be mapped to a standard problem with the following transformation

 \beq p \ = \ \frac{i}{\sqrt{\beta}} \ \tanh ( \alpha \chi ) \ \ \ \ \ \ \alpha = \sqrt{\beta} \ m \ \omega
 \label{53} \eeq

from which the resulting Lagrangian has a well known form

 \beq {\cal L} \ \sim \frac{m}{2} \ \dot{\chi}^2  \ + \ \frac{1}{2m \beta \cosh^2 (  \alpha \ \chi ) } \label{54} \eeq

We have been able to map the non-commutative quantization formula to a classical problem of quantum mechanics with  potential

\beq V(\chi) \ = \ - \ \frac{1}{2m \beta \cosh^2 (  \alpha \ \chi ) } \label{55} \eeq

The connection between the non-commutative quantization of an harmonic oscillator and this precise problem of ordinary quantum mechanics has been already noticed in our previous paper \cite{5}. Now we recover it with the path integral formalism, giving us stronger confidence on the formal manipulations made in the third section.

\section{Conclusions}

The introduction of a fundamental minimal length in a quantum system gives rise to important consequences that are still mostly unexplored. For example a simple lesson that we learn from this article is that a Lagrangian formalism is possible in presence of modified Poisson brackets of a Hamiltonian system. Since the path integral formalism is strictly connected to the Lagrangian formalism, this implies that the non-commutative quantization of a $1d$ non-relativistic system with generic potential $V(x)$ can be represented through a functional integration.
All these results have been rigorously tested both at a classical level with the equations of motion and at a quantum level with the non-commutative quantization of a harmonic oscillator.

It is clear that non-commutativity leads to more complicated formulas and unfortunately most of the potentials $V(x)$ are non integrable systems, therefore our path integral representation should be investigated through perturbative methods. Another development could be the generalization to higher degrees of freedom, with the introduction of the Snyder geometry. We expect that a deeper study is needed to overcome these problems.


\begin{thebibliography}{999}
\bibitem{1} D. J. Gross, P. F. Mende, Nucl. Phys. {\bf B303} (1988) 407.
\bibitem{2} G. Mangano, J. Math. Phys. {\bf 39} (1998) 2584; gr-qc/9705040v2.
\bibitem{3} A. Kempf, G. Mangano and R. B. Mann, Phys. Rev. D {\bf 52} (1995) 1108.
\bibitem{4} S. Mignemi, arXiv:1308.0673v2 [hep-th] (2013).
\bibitem{5} P. Valtancoli, Mod. Phys. Lett. A {\bf 27} (2012) 1250107.

\end{thebibliography}
\end{document}